\documentclass[epjST]{svjour}

\usepackage{graphics,graphicx,dcolumn,bm,fleqn,epic,eepic,float,epsfig}
\usepackage{amssymb,amsmath,multirow,rotate,color,float}
\usepackage{epstopdf}
\usepackage{times}
\usepackage{color}
\usepackage{soul}                    % package needed for overstriking
\definecolor{red}{rgb}{1,0,0}
\definecolor{green}{rgb}{0,1,0}
\definecolor{blue}{rgb}{0,0,1}

\begin{document}

\title{From human mobility to renewable energies}

%\title{Big DATA: an opportunity to assess scale phenomena}

\subtitle{Big data analysis to approach worldwide multiscale phenomena}

\author{Frank Raischel\inst{1,2} \and
        Adriano Moreira\inst{3}  \and
        Pedro G.~Lind\inst{4}}

%\institute{Insert the first address here \and the second here \and ...}
\institute{Instituto Dom Luiz, CGUL, University of Lisbon, %1749-016 Lisbon, 
           Portugal \and
           Department of Theoretical Physics, University of Debrecen, 
           Debrecen, Hungary \and
           Algoritmi Research Centre, Universidade do Minho,
           %Campus de Azur\'em, 4800-058 
           Guimar\~aes, Portugal \and
           ForWind and TWIST, Institute of Physics, %Carl-von-Ossietzky 
           University of Oldenburg, %DE-26111 Oldenburg, 
           Germany}

%\date{\today}

%%%%%ABSTRACT
\abstract{%
We address and discuss recent trends in the analysis of big data sets,
with the emphasis on studying multiscale phenomena.
Applications of big data analysis
in different scientific fields are described and 
two particular examples of multiscale phenomena are explored in
more detail.
The first one deals with wind power production at the
scale of single wind turbines, the scale of entire wind farms and
also at the scale of a whole country.
Using open source data 
we show that the wind power production has an intermittent character
at all those three scales, 
with implications for defining
adequate strategies for stable energy production.
The second example concerns the dynamics underlying human mobility, 
which presents different features at different scales.
For that end, we analyze $12$-month data of the Eduroam
database within Portuguese universities, and find that,
at the smallest scales, typically 
within a set of a few adjacent buildings, the characteristic exponents
of average displacements are different from the ones found at the scale
of one country or one continent.
}

%%%%PACS e Keywords
%\pacs{%
%%      89.75.Kd, %Pattern formation in complex systems, 
%%      01.75.+m, %Science and society, 
%%      89.90.+n} %Interdisciplinary physics, see section new topics in, 
%       89.75.Da, %Systems obeying scaling laws
%       89.75.Fb, %Structures and organization in complex systems
%       89.75.Hc} %Networks and genealogical trees 

%%02.50.Ga,  %Markov processes
%%      02.50.Ey,  %Stochastic processes
%%      92.70.Gt}   %Climate dynamics
%%            92.60.Sz \sep   % Air pollution

%\keywords{Networks, Data Analysis, Human mobility}

\maketitle

%\tableofcontents

%%%%%%%%%%%%%%%%%%%%%%%%%%%%%%%%%%%%%%%%%%%%%%%%%%%%%%%%%%%%%%%%%%%%%%
%%%%%%%%%%%%%%%%%%%%% TEXT %%%%%%%%%%%%%%%%%%%%%%%%%%%%%%%%%%%%%%%%%%%
%%%%%%%%%%%%%%%%%%%%%%%%%%%%%%%%%%%%%%%%%%%%%%%%%%%%%%%%%%%%%%%%%%%%%%

%%%%%%%%%%%%%%%%%
\section{Big data: the emergence of a new paradigm}

In some sense, the notion of Information Age is slowly 
fading from the perception of our society.
For decades, national and international research has
been creating impressive amounts of data, and often from various 
unrelated measurement sources\cite{CERN}. 
Computational methods have been serving the needs of the scientific 
community and the need for higher computational power has driven the 
invention of more efficient computational codes and triggered the 
development  of new hardware. 
Simultaneously, the internet has been instrumental in fostering 
international research collaborations and divulging scientific 
knowledge.

Today, obtaining data is no longer a problem. 
Information is there, available for everyone at any time. 
Given the computational power of today's computers and clusters, 
even single research groups can create and store large data volumes.
The challenge in today's research is more often what to do with the
data we have. How to manage all the information we have in such large 
data sources? Which phenomena can we now study?
The recent technological progress together with the new challenges 
naturally lead to a new paradigm in computational sciences, which is 
partially described by the term {\it big data}. 

In this paper we discuss the utility of big data for approaching the
empirical study of phenomena up to now unreachable.
We discuss the usage of big data in the study of two 
specific phenomena, that up to now were unreachable,
namely wind energy production and human 
mobility. These phenomena cover several spatial scales coupled with 
each other, ranging from a few meters up to several hundreds of 
kilometers.
Such phenomena are usually called multiscale phenomena.
%%%%%%%%%%%%%%%%%%%%%%%%%%%%%%%%%%%%%%%%%%%%%%%%%%%%%%%%%%%%%%%%%%%%%%%
\begin{figure}[t]
\centering
\includegraphics[width=0.95\textwidth]{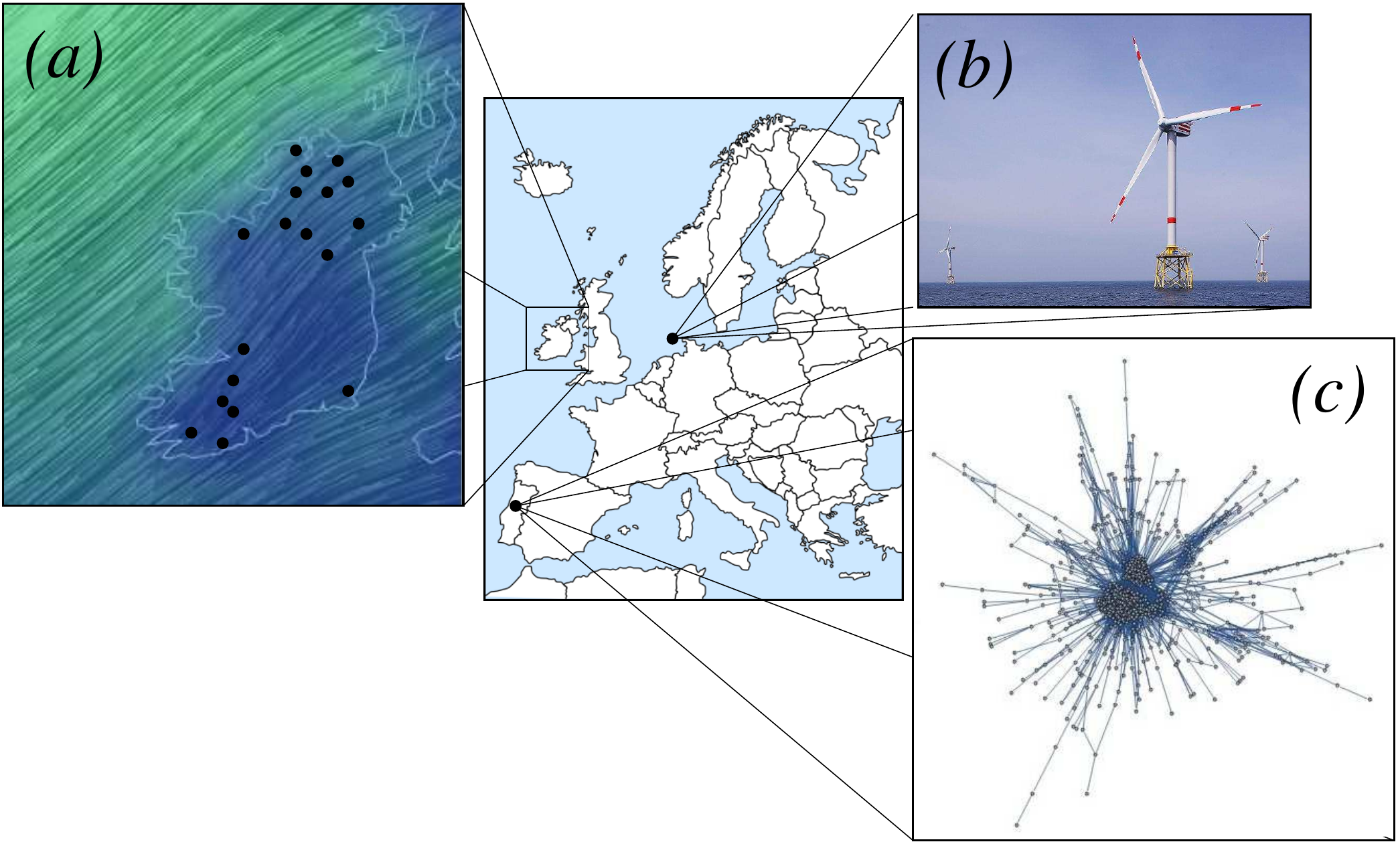}
\caption{\protect (Color online)
         Big data sets and their application to multiscale phenomena:
         wind energy production at the scale of
         {\bf (a)} one wind farm or an entire country and of
         {\bf (b)} single wind turbines.
         In {\bf (c)} human mobility, where one shows a subset
         of the Eduroam network 
         comprehending the data from University of Minho
         (January-December 2011), with its 
         workers and students spreaded throughout other Eduroam 
         networks in Portugal.}
\label{fig1}
\end{figure}
%%%%%%%%%%%%%%%%%%%%%%%%%%%%%%%%%%%%%%%%%%%%%%%%%%%%%%%%%%%%%%%%%%%%%%%

Big data is generally understood as notion of massive amounts of data, 
often of social and economical origin and/or relevance, whose storage, 
loading into memory, or computational processing by far exceeds the 
computational power of a single PC, and therefore requires the use of 
parallel file systems and processors and the development of the 
corresponding parallelized software, not only for algorithmic
computation, 
but also for storing and accessing the data, see
\cite{bigdata} about the concept and \cite{nelson2008} about how
it existed even before computers.
The new big data paradigm is therefore rooted in four technological 
and societal developments.

First, 
rapid and low-cost electronics for sensoring and data acquisition
allow us to obtain the so-called {\it rise of electronics}: modern and low-cost 
electronic data acquisition and sensoring devices give  the  technological 
possibility to acquire in real time large-scale 
experimental, simulational or social data.  
Important examples are, of course, satellites, cellular phones,
D/A converters, and the internet, which itself is a considerable source
of social data.
% as well as multichannel data acquisition systems and
%any sort of data found on the internet. 

Second, we now achieve data completeness on scales.
Among many others one may refer to the example 
of travel and cell phone data, as well as health data, 
data from electronic payment systems or high-frequency 
trading data.
%, multiscale material models 
%and simulations of complex liquids, or even financial data from 
%high-frequency trading.
In geophysics, meteorological data cover geographically scales
as small as a few kilometers up to the entire global scale.
When referring to social data the complete demographic
databases may be available. In physical phenomena, all relevant scales 
of physical multiscale problems such as turbulence, multiscale material
problems or complex fluids can nowadays possible
to be empirically assessed.

Third, it is also necessary to assure {\it data availability} as well
as computational power for developing {\it data mining procedures}.
This means that we have on the one hand the technological, organizational, 
economical and efficient means to centrally  store, share and transfer   
the data.
On the other hand, the practical usefullness of the stored data is strongly
influenced by the technological means to 
efficiently and rapidly find patterns in these data.

Finally, one should have a storage and mining system that enables
the {\it combination of sources}. 
%The possibility to combine and cross data from various 
Today's technology enables the combination of the 
aforementioned sources, e.g.~location and social interaction from 
cell phone location and connection data, 
travel  and spending patterns 
%(from cell phone location and electronic payment data), 
or the 
combination of meteorological data, wind power generation and 
electrical network activity.

These mainly tehnological developments have had a strong impact in
physics, which has witnessed the establishment of computational method 
as a third pillar in empirical sciences, next to theory and 
experimentation\cite{historyofscience}. 
However, big data is not defined by the magnitude of the data alone, 
but rather by the statistical complexity of the problem that is being 
addressed.
Of particular interest
are multiscale phenomena, since they require typically large amounts
of data from different sources.

We start in Sec.~\ref{sec:other} by describing in some detail
applications of big data for various multiscale phenomena.
In Sec.~\ref{sec:wind} we treat the specific case of big data analysis
in wind energy research, 
comparing the power increments found in the full wind power production
data in Ireland with those from a large park in Australia and a single
turbine in 
%analyzing data sets 
%containing the full power production 
%in Irish and Australian wind farms (Fig.~\ref{fig1}a), comparing
%the power increments with the ones observed in data series of power 
%production in single wind turbines in 
an off-shore wind farm in Germany's North Sea (Fig.~\ref{fig1}b).
In Sec.~\ref{sec:human} we approach the problem of the patterns
describing human mobility, describing recent results at large scales
and showing how they differ from human motion at the smallest scales.
Section \ref{sec:conclusions} concludes the paper.

%%%%%%%%%%%%%%%
\section{Big Data for World-wide and Multiscale Problems}
%            in computational physics}
\label{sec:other}

One important aspect in big data is that it enabled physicists to 
finally treat 
physical and statistical problems not only in a crude approximation, 
on a single relevant scale, on a coarse grid, or for a small sample of data, 
but on a full set of different scales, either in time, in space or in 
their organizational nature.
In this section we discuss briefly some examples of world-wide 
multiscale data.

One phenomena that can now be approached refers to 
the entire global data sets for human travel patterns, 
or social data sets representing entire populations.
It is known\cite{gonzaleznature,brockmannnature} that human
mobility follows non-Brownian laws when observed at large 
scales, such as that of an entire country. 
However, as we show below, for the smallest scales, 
comprehending streets and few buildings, we find human motion
to be almost Brownian\cite{adriano}.

The international financial markets\cite{econophysics} are another example 
of an extremely interlinked system, with information flow on various 
time scales, an effect that is ever increasing due to globalization 
and high speed trading\cite{Kenett2013}.
It is now well accepted that a realistic description of financial 
markets can be achieved through agent models, which try to recreate 
financial interdependencies among a large number of  individuals which 
differ by behavior and available assets\cite{levy_agents}. 
These models allow notably for the description of 
failure avalanches\cite{cruz_financial}.
An interesting example of the information exchange  between different 
world-wide multiscale  systems is the predictive character  of 
human internet searches on stock market trades\cite{Preis2013}. Even 
when considering only small samples of data sets representing connected 
agents, analyzing the correlations between them quickly leads to 
complex problems\cite{Laloux1999,plerou_cor}.
Further examples range from uncovering the
correct statistical description of financial time 
series\cite{Camargo2013}, as far as they exist in a non-stationary 
system\cite{mccauley2004}, to devising trading strategies which could 
mitigate sudden crashes\cite{Biondo2013}, to finding regulatory measures
which can stabilize the markets\cite{cruz_financial}.

Nevertheless, an important note is in order here: financial
markets are not physical systems. Rather, the principal driving force of 
markets is profit, where profit is to be gained by -- at least partially 
irrational -- individuals competing for advantages in information, and
influencing each others' opinions\cite{Kinzel2003,sornette2009}. 
This reaction to new information leads to ``reclective''
feedback loops\cite{sornette2012} and is in strong contrast to the 
notion of invariant natural laws found in the sciences. 
As Feynman said fittingly
``Imagine how much harder physics would be if electrons had 
feelings!'', cited in \cite{Lo2010}.

Another important field where processes in scale are abundant
is geophysics.
Here, 
energy flows across structures spanning several orders of magnitude.
Earthquakes, which are essentially a fracturing  phenomenon, show a 
flow of energy release events  from the micro- to the macroscale. 
Present-day state of the art is the microscopic simulation of fracture 
on the molecular scale\cite{Abraham2003}, 
real-time observation at the nanoscale\cite{prades2005},
the stochastic or continuum 
description at the  mesoscale\cite{kun2006,kun2014}, and the 
statistical description of earthquakes on the fault system or global 
scale\cite{sornette2008,Holliday2008}.

Recent advantages in the modeling of fracture of heterogeneous materials 
could give an answer about the nature of the fracture phase 
transition\cite{Shekhawat2013}. 
At present, it remains an unresolved problem to correctly understand the 
details of damage propagation in models of earthquakes  and experimental 
data, e.g.~the statistics of intershock 
sequences\cite{Bottiglieri2010,Touati2009},  and the recurrence of events
and record breakings\cite{Davidsen2008}.
The possibility of predicting individual  earthquakes remains, however, 
a subject with an unclear 
outlook\cite{pradhan2005,Lippiello2012,Main1999,Main2012}.  

Even on the geological time- and length scales, models exist to describe 
the formation and phenomenology of geomorphological features, such as 
river deltas\cite{Seybold2007}, coastal dunes\cite{duran2013},
watersheds\cite{Fehr2011}, or coastlines\cite{morais2011}.
The latter two models are based on scaling processes related to  
percolation and represent long-scale correlations in the formation 
process. 

Similar considerations lead to models of porous rock 
formations\cite{Biswal2007}, which often carry fossil resources, 
leading to the question of how these reservoirs can be exploited most 
efficiently\cite{Schrenk2012}. Here simulations of the complex 
fluid-fluid and fluid-rock interfaces are also needed  to gain further  
insight\cite{Succi2001a,narvaez2010,narvaez2013}.  

To end this overview of research topics, there is also the
problem of wind energy production and its fluctuations 
in time. Wind energy production is essentially determined by large-scale 
turbulent atmospheric dynamics 
and extreme event statistics, two topics that still require a 
multidisciplinary
comprehensive approach, from both the 
earth sciences \cite{Lovejoy2009,Donner2009a} 
and physics\cite{Friedrich2011a}.
 
Recently, new insight has been found into how intermittency 
in wind energy production is preserved  when going from the scale of 
single wind turbines to the scale of entire wind farms and even to 
the scale of countries having several wind 
farms\cite{rezassub,pats,oli}.  
This has raised new questions on  how to stabilize energy production 
on the scale of large economies.
Furthermore, the role of  long-range correlations in wind velocities  
due to meteorological systems on the synoptic scale remains an open 
question\cite{vitor_wr}.

As one can see, we find world-wide multiscale processes spanning 
from social systems, through biology, and geological processes. 
In the following, we will take a closer look at two of these 
phenomena.
The first one, world-wide wind-power generation, not only is 
affected by large-scale meteorological processes, but then
it also links the produced power to continental electric grids.
The second phenomenon is 
human mobility, and its importance is intrinsically linked to multiscale 
multiscale technological applications, 
e.g.~flight management and cell phone networks,
as well as to biology, when addressing the world-wide spreading of 
diseases.

%%%%%%%%%%%%%%%%%%%%%%%%%%%%%%%%%%%%%%%
\section{Big data in geophysics: the example of wind 
energy}
\label{sec:wind}

Wind energy is one of the best candidates to answer the world-wide 
energetic problem\cite{johnsonWindBook}, and simultaneously it is a
``clean'' source of energy. 
However it has a drawback: it reflects the turbulent nature of 
the wind itself. 
As it is known\cite{windenergyhandbook}, since wind speed presents 
non-Gaussian fluctuations in time, the power output of one turbine
also shows this intermittent behavior\cite{patrickprl} making 
predictions of energy production rather difficult.
Moreover, 
power grids fed only from wind energy alone would be unstable, with 
periods of large production alternated with periods of very low 
production.
These facts raise
challenges to the wind power producers who need 
to trade on these variations in the electricity markets\cite{teresaspaper}.
Large fluctuations in the energy produced can impede the
matching to the committed energy.

\begin{figure}[t]
\centering
\includegraphics[width=0.95\textwidth]{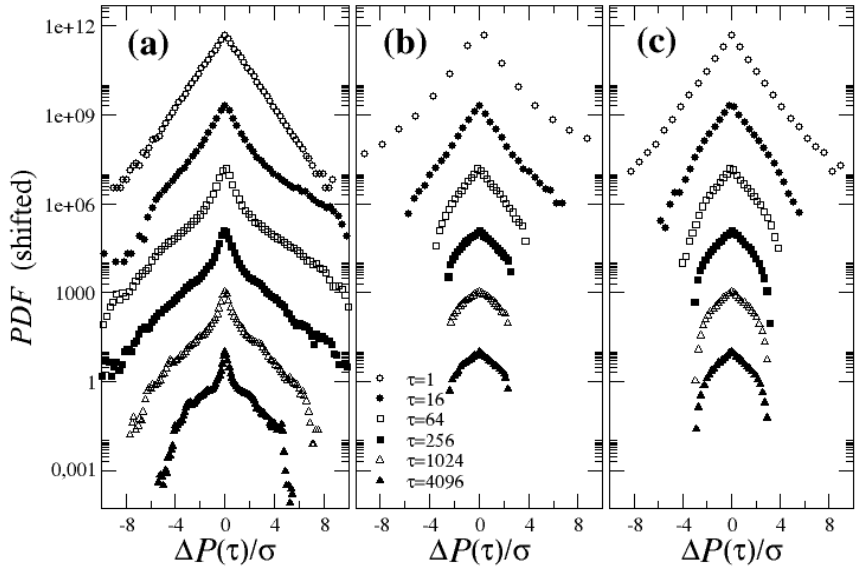}
\caption{\protect 
         Big data sets and its application to multiscale phenomena:
         wind energy production at the scale of
         {\bf (a)} single wind turbines (in units of seconds), of 
         {\bf (b)} a large wind farm in Australia (in units
                   of $5$ minutes) and of 
         {\bf (c)} an entire country, Ireland, 
                   for which a detrending
         procedure was applied as described in \cite{oli} 
         (in units of $15$ minutes).
         In all plots PDFs are shifted vertically for better
         visualization.}
\label{fig2}
\end{figure}

These large fluctuations are a function of the time scale of the
observation and disappear for large enough time increments.
%When such large statistical fluctuations occur, one way to overwhelm
%them would be to consider fluctuations at larger time steps.
Indeed, similar to what happens in the stock 
market\cite{naturepeinke}, while the increments of power production
evidence significant deviations from the Gaussian regime with
heavy tails, for large time increments these tails disappear.
Figure \ref{fig2}a shows the increment statistics
of the power output defined as
\begin{equation}
\Delta P_{\tau}(t) = P(t+\tau)-P(t) ,
\label{increment}
\end{equation}
where $P(t)$ is the power output measured at time-step $t$ and
$\tau$ is a fixed time-gap between two measures.
As we see in Fig.~\ref{fig2}a the heavy-tails disappear as the 
time-gap increases from one second up to more than one hour.

However, expected returns in power production 
must sometimes be taken within a time period shorter than one
hour, increasing the risk associated with the expected values.
Recently, some variants of the standard risk-return approach
have been proposed\cite{teresaspaper} for solving this shortcoming.
%%%%%%%%%%%%%%%%%%%%%%%%%%%%%%%%%%%%%%%%%%%%%%%%%%%%%%%%%%%%%%%%%%%%%%
%%%%%%%%%%%%%%%%%%%%%%%%%%%%%%%%%%%%%%%%%%%%%%%%%%%%%%%%%%%%%%%%%%%%%%

The single turbine data was provided by Senvion\cite{data} 
and comprehends the 
full month of January 2013, measured at one wind turbine of the Alpha 
Ventus wind farm in Germany, located approximately at 
$54.3^o$N-$6.5^o$W.

Another way to ``smoothen out'' the impact of these short-term
fluctuations could be to add more turbines to a given network.
If a single turbine would show Gaussian statistics in its power 
output, even if they were not completely uncorrelated, then, 
according to the central limit theorem, the heavy-tails of the 
distribution should disappear as one considers the joint power 
output of more and more turbines.

However what we find in the case of wind
energy production is an intermittent behavior at all three scales.
Figure \ref{fig2}b and \ref{fig2}c show the distribution of
power increments,
aggregating data of an entire wind farm in Australia and
of all wind farms in Ireland, respectively.
As one sees in Fig.~\ref{fig2} the large fluctuations of energy 
production at the level of one single turbine are also observed
on the power production of an entire wind farm
or on the sum of all wind farms within one country.
Similar results were observed in previous works\cite{pats,oli}.

One possible cause for the emergence of heavy tails at several 
spatial scales is the existence of uncorrelated, 
stable-L\'evy distributed sources. 
Another cause is the existence of correlations between the wind 
conditions at all these scales.
In the case of Ireland, the wind flux streams 
drawn in Fig.~\ref{fig1}a clearly sweep the entire country in
a similar way. In other words, Ireland lies typically in the 
same weather system and therefore the wind shows the same features
whether it is observed at one single turbine, in an entire wind 
farm or as a pattern in a full country.

By taking adjacent countries comprising an area larger than
the entire weather system, wind turbines and wind farms in
that larger area will be differently affected by the weather
system and therefore will eventually lead to uncorrelated 
wind conditions. Consequently, having ``uncorrelated''
wind turbines and wind farms opens the possibility for more
stable power grids and a better matching between produced 
and committed energy.
Moreover, recently, it has been shown that combining different
renewable energy sources, such as wind and solar sources,
would allow to obtain a more constant combined power 
production rate\cite{rezassub}.

%%%%%%%%%%%%%
\section{Big data in society: the example of human mobility}
\label{sec:human}

Understanding human motion from small scales, such as buildings and
streets up to larger ones comprising cities, countries and continents,
is fundamental for a variety of topics such
as spreading of diseases, optimization of telecommunication networks, 
urban planning, and tourism management.
Several groups have been modeling human 
motion\cite{gonzaleznature,brockmannnature,ahas2006,ahas2007},
showing that human traveling distances within the USA decay 
as a power-law\cite{brockmannnature}, and that there is one single 
probability distribution for time returns to previous 
locations\cite{gonzaleznature}.

In Ref.~\cite{brockmannnature} the authors use
data obtained from an online tracker system where registered 
users reported the observation of marked US dollars bills,
while in Ref.~\cite{gonzaleznature} one data set is used,
containing positioning records of around $10^5$ users of 
a cellular network.
While these data sets were important to uncover mobility 
patterns at the level of the USA, 
two drawbacks of the data sets used so far should be stressed.
First, in both cases the data records correspond to positions 
collected at sparse time steps and no continuous tracking is 
possible.
Second, the accuracy of position measurement were constrained
to e.g.~the cell ID with a coverage area over several square 
kilometers.
Therefore, human mobility at smaller scales still lacks to be 
addressed, mostly because of the nature of available data.

In this section we analyze a large data set of individual 
locations in a continuous time tracking system at scales
as small as buildings and city streets.
The data set we analyze is extracted from the Eduroam networks 
at Portuguese Universities (see Fig.~\ref{fig1}c), a part of
a large data set comprising entire Europe.
By analyzing the Eduroam net at the level of one single university
we achieve the smallest scale at which human motion takes
place, as addressed in 
Refs.~\cite{adriano,adriano1,adriano2,adriano3}.

The Eduroam data set comprises one year of collected 
information with a sample frequency of one second, a time 
step that enables to assume the data as continuous monitoring 
of human motion.
The spatial resolution is of the order of a few meters, the
typical distance between adjacent access points (APs).
At these small scales one expects that human trajectories
are closer to the Brownian regime than what is observed at
the larger spatial scales reported in 
Refs.~\cite{gonzaleznature,brockmannnature}.
We will show that indeed this is the case.

The data set comprehends an Eduroam RADIUS log file from the 
University of Minho (Portugal) infrastructure, including the 
full year of 2011.
The file contains a total of $15,892,009$ data records
$7,937,245$ of them refer to ``start-events'' and $7,954,764$ 
refer to ``stop-events''. 
A few of these records ($0.65\%$) have been filtered, since they 
are incomplete or do not represent unique Access Points or unique Stations. 
After filtering, we obtain $7,902,828$ records for processing and 
analysis.

First, we extract the topological structure of the
network of APs. More than the spatial distances between
APs joined by paths such as stairs, corridors or open
spaces, we deal with a topological distance between
APs that reflects how often people tend to move between
each pair of APs. 
Such an approach in constructing a complex network 
has been applied frequently to address social and 
environmental 
problems\cite{lind05,gonzalez06,gonzalez06b,lind07,lind07b}.
Still, we introduce here a novel procedure for extracting 
the network on which persons move directly from empirical 
data. Our approach can be easily extended to the full
European Eduroam data set.

Using all the records at University of Minho we construct 
a weighted AP network considering all $723$ APs detected during
each full month of 2011. 

For extracting the AP network, we consider the Eduroam 
database as an output file with three single fields 
for each register, namely the user $i$ (or alternatively 
the device one user is using when connected to the net),
the AP $I$, where the user is connected to and
the time $t$ in units of the time step $\Delta t=1$ second, 
which is typically the size of
the smallest time lag between successive registers.
Henceforth, small letters indicate properties of the
users, while capital letters indicate properties of
the APs.

We start by counting the total
number of connected users to one 
particular AP $I$ at time step $t$, yielding the AP fitness 
$G_I(t)$. 
At each time-step $t$, $G_I(t)$ at AP $I$ equals the total 
number of associations occurred at $I$ minus the total number 
of disassociations, occurred within the full time span from 
the beginning of the observation period up to time $t$.
The full time span considered is of $24$ hours, beyond which
we assume that each user cannot be further connected.
Our results have shown that neglecting sessions longer than 
24 hours has no significant impact in the following 
analysis\cite{adriano}.

%Displacements between APs, defined as the disassociation from 
%one AP $I$ and subsequent association with AP $J$, exhibit a periodic 
%behavior as illustrated by the distribution of the corresponding 
%time spans (see Fig.~\ref{fig3}c). However, the majority of the 
%displacements have a short duration, representing fast displacements 
%among nearby APs.
%%%%%%%%%%%%%%%%%%%%%%%%%%%%%%%%%%%%%%%%%%%%%%%%%%%%%%%%%%%%%%%%%%%%%%%
\begin{figure}[t]
\centering
\includegraphics[width=0.33\textwidth]{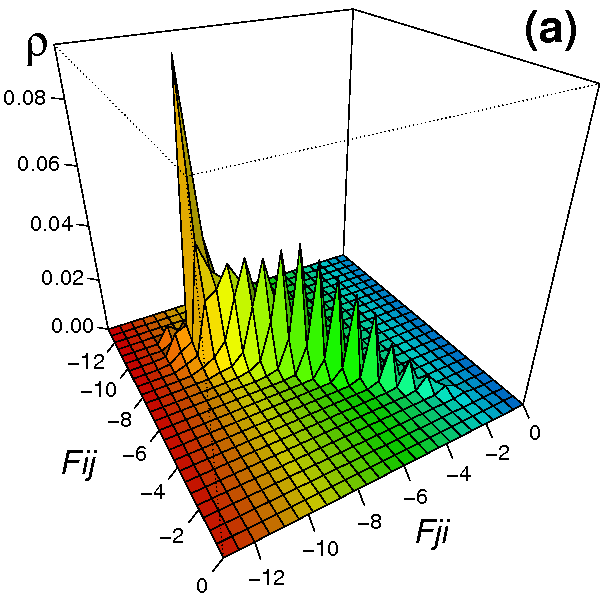}%
\includegraphics[width=0.66\textwidth]{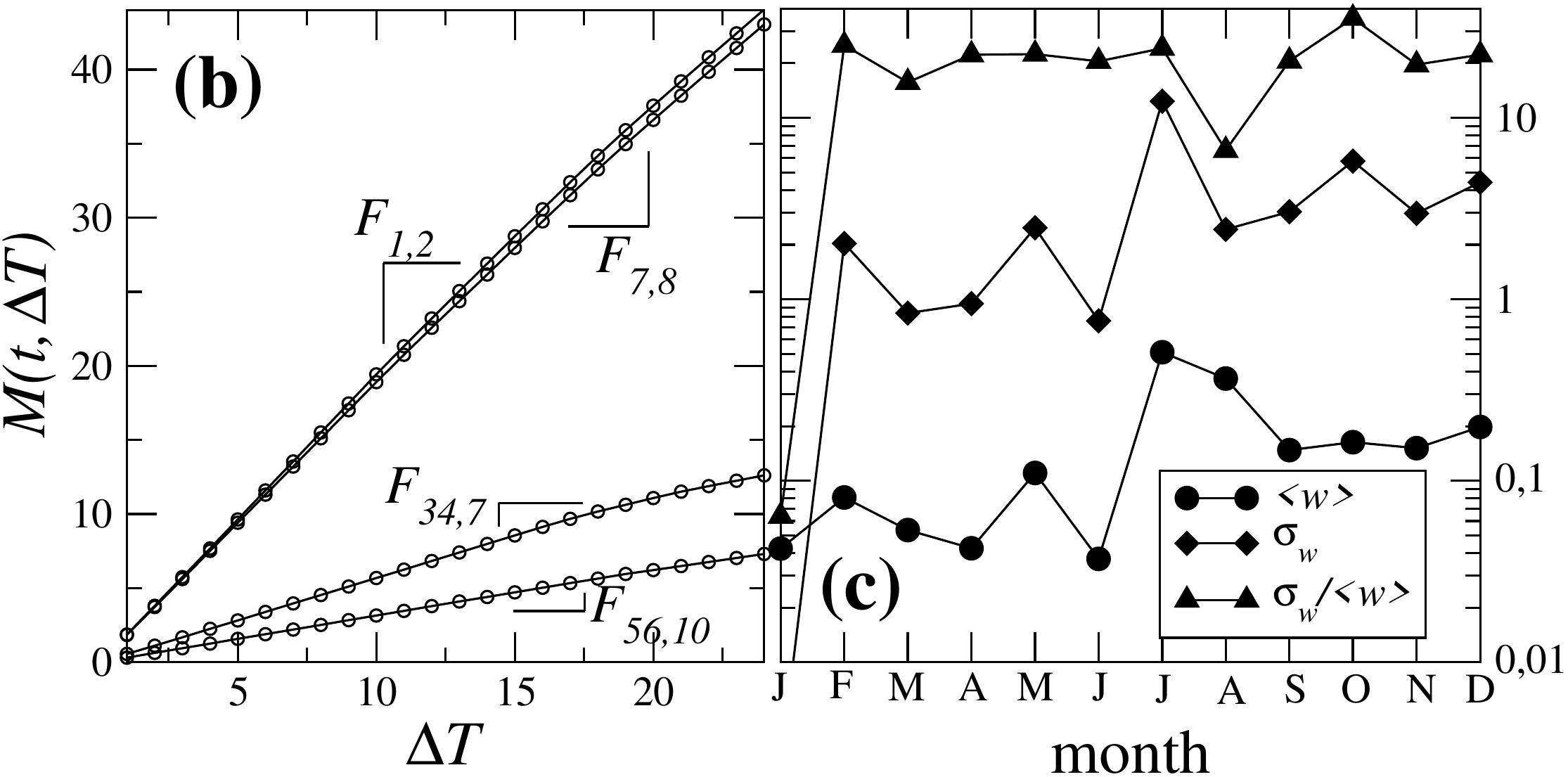}
\caption{\protect 
   {\bf (a)} Histogram of the average flux between each 
   pair of APs in both directions, showing a symmetry relation 
   $\langle F_{IJ}\rangle \simeq \langle F_{JI}\rangle$ 
   which supports the definition for the (symmetric) weights 
   $W_{IJ}$ in the AP network (see text).
   {\bf (b)} Number $M$ of AP switches as a function of the time 
   window $\Delta T$, defining the flux 
   $F_{IJ} =dM/d(\Delta T)\sim M/\Delta T$ 
   (see Eq.~(\ref{instflux})) which is approximately 
   constant for $\Delta T$.
   Shown is the slope for four different pairs $IJ$ of APs taken at 
   the first day of 2011. 
   {\bf (c)} Weight average $\langle W\rangle$ and standard deviation 
   $\sigma_W$ as a function of month, using January as the reference 
   month.}
\label{fig3}
\end{figure}
%%%%%%%%%%%%%%%%%%%%%%%%%%%%%%%%%%%%%%%%%%%%%%%%%%%%%%%%%%%%%%%%%%%%%%%

The network of APs has a weight matrix defined by the 
weights $W_{IJ}=1/D_{IJ}$, where $D_{IJ}=D_{JI}$ is a proper topological 
distance between APs $I$ and $J$.
Since we do not have the location of the APs and we do not know
exactly the full number of constraints that condition the displacement
from one AP to the next one, we introduce a topological distance as
follows.

First, we compare
the average fitness $\langle G_I(t) \rangle$ at 
each AP $I$ and the instantaneous flux $F_{IJ}$ between $I$ and $J$ 
defined as 
\begin{equation}
F_{IJ}(t)=\frac{dM_{IJ}(t,\Delta T)}{d(\Delta T)} ,
\label{instflux} 
\end{equation}
where $M_{IJ}(t,\Delta T)$
is the number of users that associate to AP $J$ after 
disassociating from AP $I$, within a time window 
$\Delta T \lesssim 1$ day start at time $t$.
Figure \ref{fig3}a shows the frequency of each flux strength
$F_{IJ}$, in one direction, and the flux $F_{JI}$, in the opposite 
direction, showing a symmetry, $ F_{IJ} \sim F_{JI} $.
As shown in Fig.~\ref{fig3}b, for four different pairs $IJ$ of APs, 
the average flux is approximately constant within time windows up to 
one day. 
Therefore, to reduce the computation time,
we consider henceforth the average flux within $\Delta T=4$ 
hours as an estimate of the flux between each pair of APs to reduce 
the computation time.

From these quantities, $\langle F_{IJ}\rangle$ and 
$\langle G_{I}(t)\rangle$ one finally defines the weight of each
connection as 
\begin{equation}
W_{IJ}\equiv \frac{1}{D_{IJ}} = \tfrac{1}{2} \left (
       \frac{\langle  F_{IJ}(t)\rangle}{\langle  G_{I}(t)\rangle}+
       \frac{\langle  F_{JI}(t)\rangle}{\langle  G_{J}(t)\rangle}
     \right ) .   
\label{eqFlux}
\end{equation}
From such a definition the weights are properly averaged over their
fitness and simultaneously contain the same symmetry as the fluxes,
$W_{IJ} = W_{JI}$ (see Fig.~\ref{fig3}a).

Since there are seasonal fluctuations from month to month, as
shown in Fig.~\ref{fig3}c, the average fitness at each AP is 
considered for a time window of one month and repeated for every 
month during the entire year. 
A similar approach has been used for the calculation of the average 
flux, which yields one AP network for each month.
The weight average $\langle W \rangle$ -- over all AP connections $IJ$ -- 
varies from month to month as well as the corresponding deviations 
\begin{equation}
\sigma_W(t) = \left ( \frac{1}{N-1}\sum_{connections} 
\left ( W_{IJ}(t)-\langle W\rangle (0)\right )^2\right )^{1/2}
\end{equation}
from a reference month average, chosen to be January.
While some significant variations are observed during the summer 
vacations, the normalized deviations $\sigma_W/\langle W\rangle$ are 
approximately constant.
This result also highlights the temporal evolution on the usage of the 
network, with more and more users accessing the network and also 
more APs being deployed to improve the network capacity.
%%%%%%%%%%%%%%%%%%%%%%%%%%%%%%%%%%%%%%%%%%%%%%%%%%%%%%%%%%%%%%%%%%%%%%%
\begin{figure}[t]
   \centering
   \includegraphics[width=0.7\textwidth]{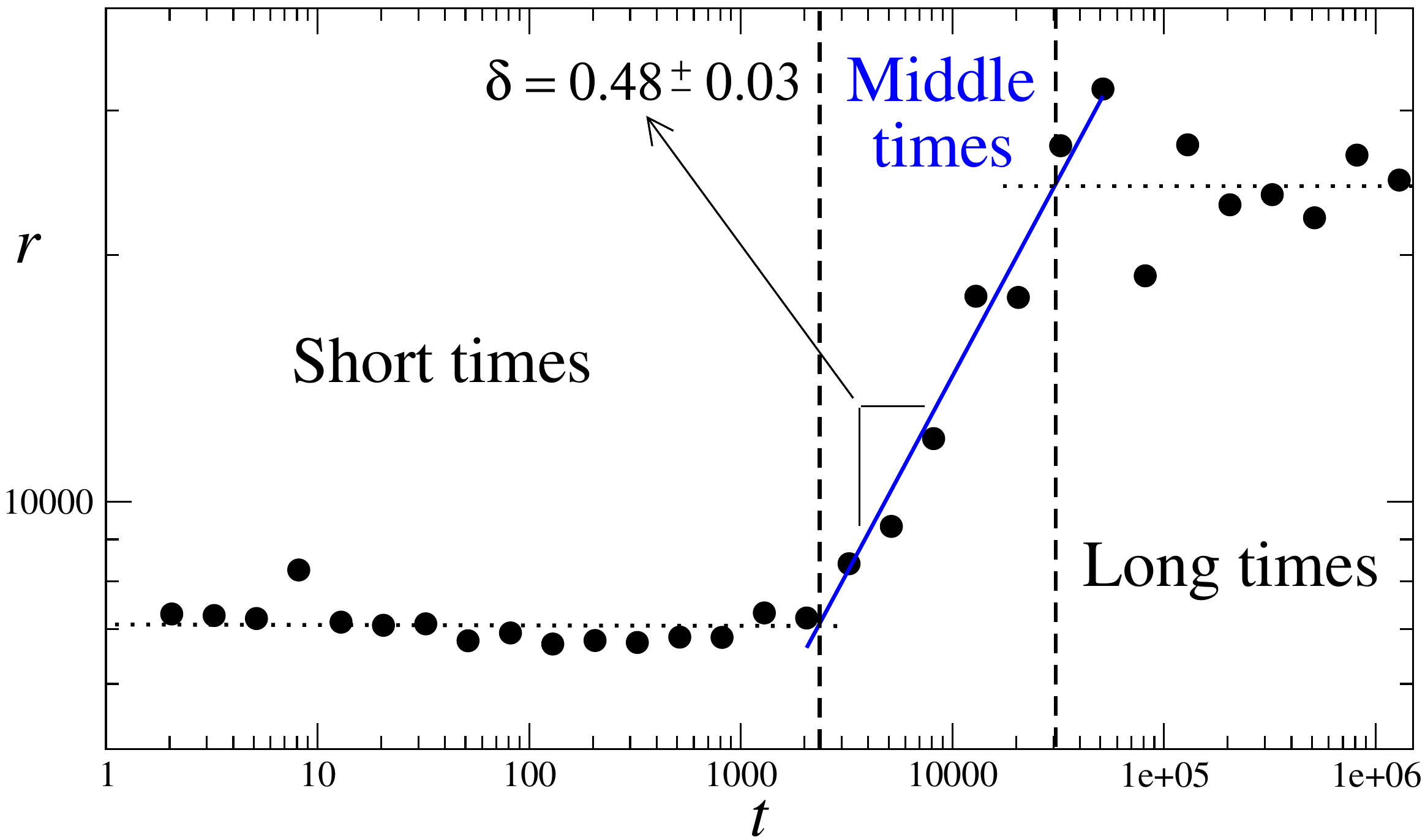}
   \caption{\protect 
   Average shortest distance $r(t)$ of human trajectories within 
   Portuguese Universities as a function of time (in seconds). 
   While for short times people tend to remain at the same place 
   (connected to the same AP), with $r_{min}$, beyond around $20$ 
   minutes there is a marked scaling of the displacement with time, 
   $r\sim t^\delta$ with $\delta \sim 0.5$, describing the normal 
   Brownian diffusion regime.}
\label{fig4}
\end{figure}
%%%%%%%%%%%%%%%%%%%%%%%%%%%%%%%%%%%%%%%%%%%%%%%%%%%%%%%%%%%%%%%%%%%%%%%

%%%%%%%%%
%\subsection{Mobility patterns at small scales}
%\label{sec:mobility}

%To study the mobility within the universities we make use of the 
%inverse weight of the connections between adjacent APs. Since the 
%weight measures the normalized flux between two adjacent APs, say 
%$I$ and $J$, its inverse value can be taken as a good estimate of 
%the topological distance $D_{IJ}$ between those APs, which of course 
%defines a symmetric matrix ($D_{IJ}=D_{JI}$).

Having defined the weight, one can immediately obtain the 
matrix of topological distances $D_{IJ}=1/W_{IJ}$ between APs. 
We symbolize the position of one individual $i$ at time $t$ 
through its trajectory as
$r_i(t)$ with reference to an initial position, labeled AP $I_0$.
In other words, if person $i$ at time $t$ is connected to
AP $I$, its position is $D_{I_0I}$ (see Eq.~(\ref{eqFlux})
and Eq.~(\ref{eq:r}) below).
For practical purposes we assume that one person corresponds
to one single trajectory, and thus label both trajectories 
and individuals with the same label $i$.
Each trajectory followed by one person is defined by the 
succession of APs and by the corresponding shortest distances 
$D_{I_0I}$ to the initial AP $I_0$.

Keeping track of the distance from the initial position, and
averaging over all trajectories in the data sets yields the
average shortest distance from the initial position computed
as
\begin{equation}
r(t)=\langle r_i(t) \rangle_{traj}=\langle D_{I_0I} \rangle_{traj}
=\frac{1}{N_{traj}(t)}\sum_{i=1}^{N_{traj}(t)}r_i(t),
\label{eq:r}
\end{equation}
with $i$ labeling one of the $N_{traj}(t)$ individual trajectories 
observed at each time $t$ starting at $t_0$, with steps of 
$\Delta t=1$ second.

As shown in Fig.~\ref{fig4}, while for short and long times the 
displacement tends to remain approximately constant,
for middle times $r(t)$ increases with time $t$ as a power law
$r\sim t^\delta$ with $\delta\simeq 0.48\pm 0.03$.
%Having obtained the value of $\delta$ for the average distance
%one can now proceed to access dynamical features of these human 
%trajectories at small scales.
This exponent is equal, within numerical error, to the one 
known to characterize Brownian diffusion, contrary to what is
observed for large scale mobility, as reported
in Refs.~\cite{gonzaleznature,brockmannnature}.
Seemingly, people move according to large jumps within large
time-spans and over large spatial scales, but they tend to diffuse
randomly with small jumps when trajectories are tracked in shorter 
times and over the smallest spatial scales.

%%%%%%%%%%%%%%%%%%%%%%%%%%%%%%%%%%%%%%%
\section{Discussion and conclusions}
\label{sec:conclusions}

In this paper we focus on possible applications of big data set 
analysis to empirically approach multiscale phenomena.
%, something until
%recently not possible on all scales.
Providing a brief discussion in the fields of sociology, finance,
economics, physics and geophysics, we discuss in detail
two particular phenomena, namely fluctuations in wind energy 
production and human mobility.

When assessing data of
wind energy production we observed a large intermittency 
in the increments of wind power at scales up to the country level.
This intermittent power production raises difficulties in the
energy market for determining the optimal committed power outputs.
Two ways to overcome this problem could be proposed.

One possibility is to construct power grids based in wind farms
typically distributed over more than one country, in order 
to sum less correlated power outputs.
Such possibility has two inconveniences.
First, countries within Europe should agree on long-term energy 
policies and commit to the corresponding investments.
Second, even if such an agreement is met for merging into 
a European 
joint stable power grid, several problems related to storage and
distribution in large power grid are still to be 
solved\cite{andre2013}.

Another possibility would be to join different sources of 
renewable energy, for example solar and wind energy within one
country.
This possibility was recently proposed\cite{rezassub} 
although analysis of data at larger scales might provide insight 
on how reliable these pioneering studies are.

For human mobility we considered the Eduroam data set at the 
smallest scales where individuals move along streets and between 
buildings, in this case, within the University of Minho (Portugal).
While it is known that a power-law distribution is found
for the distance from starting position when one keeps track of 
individual trajectories at large scales, at small scales we found 
that human mobility closely follows Brownian motion.
The mobility network was extracted from the empirical data directly
by measuring average fluxes of people between adjacent access points
inside the university.

Since the Eduroam data base covers the whole of Europe, a possible 
next step would be to consider the joint data base of several 
European universities. 
By properly matching the mask IDs and similar 
quantities at different universities, one would be able to keep
track of individuals throughout Europe and apply the framework
described above to data sets comprising larger and larger
areas. Since the mobility of faculty, researchers and students is becoming
higher due to European programs such as Erasmus, one
should expect sufficient statistics 
in the Eduroam databases at several spatial scales.
Consequently, one should be able to
observe the transition from Brownian to non-Brownian motion,
postulated in this paper.

%%%%%%%%%%%%%%%%%%%%%%%%%%%%%%%%%%%%%%%
\section*{Acknowledgments}

The authors thank
partial support by {\it Funda\c{c}\~ao para a Ci\^encia e a Tecnologia}
under the R\&D project PTDC/EIA-EIA/113933/2009.
FR assisted in fundamental research in the frame of 
T\'AMOP 4.2.4.A/2-11-1-2012-0001 National Excellence Program.
Elaborating and operating an inland student and researcher personal 
support system, was realised with personal support. 
The project was subsidized by the European Union and co-financed by the
European Social Fund.
% and SFRH/BPD/65427/2009 (FR).
%FR also thanks Ferenc Kun, Univ. Debrecen (Hungary), for his hospitality and 
%acknowledges support from the John von Neumann International Scholarship for 
%junior foreign teachers-researchers: {\it This research was realized in the 
%frames of the highly important ``National Excellence Program - working out 
%and operating an inland student and researcher support, identification number
%TMOP 4.2.4.A/2-11-1-2012-0001. 
%The project is realized with the help of European Union
%and Hungary subsidy and co-financing by the European Social Fund''}.
PGL thanks the German Environment Ministry as part of the
research project ``Probabilistic loads description, monitoring, and
reduction for the next generation offshore wind turbines (OWEA Loads)''
under grant number 0325577B.
Authors also thank Senvion for providing the data here analyzed.
All data series were analyzed according to all confidential protocols and
were properly masked through the normalization by their highest values.
Therefore the scientific conclusions are not affected by such data
protection requirements.

%%%%%%%%%%%%%%%%%%%%%%%%%%%%%%%%%%%%%%%%%%%%%%%%%%%%%%%%%%%%%%%%%%%%%%%%%%%
%%%%%%%%%%%%%%REFERENCIAS %%%%%%%%%%%%%%%%%%%%%%%%%%%%%%%%%%%%%%%%%%%%%%%%%
%%%%%%%%%%%%%%%%%%%%%%%%%%%%%%%%%%%%%%%%%%%%%%%%%%%%%%%%%%%%%%%%%%%%%%%%%%%

\end{document}